# On Target Detection by Quantum Radar
# (Preprint)

## Gaspare Galati, Gabriele Pavan

Tor Vergata University – Department of Electronic Engineering, Rome (Italy) and CNIT - National Inter-University Consortium for Telecommunications - Research Unit of Roma Tor Vergata

gaspare.galati@uniroma2.it, gabriele.pavan@uniroma2.it

***Abstract.*** Both Noise Radar (NR) and Quantum Radar (QR), with some alleged common features, exploit the randomness of the transmitted signal to enhance radar covertness and to reduce mutual interference. While NR has been prototypically developed and successfully tested in many environments by different organizations, the significant investments on QR seem not to be followed by practically operating "outdoor" prototypes or demonstrators. Starting from the trivial fact that radar detection depends on the energy transmitted on the target and backscattered by it, some detailed evaluations in this work show that the detection performance of all the proposed QR types in the literature are orders of magnitude below the ones of a much simpler and cheaper equivalent "classical" radar set, in particular of the NR type. Moreover, the absence of a, sometimes alleged, "Quantum radar cross section" different from the radar cross section is explained. Hence, the various Quantum Radar proposals cannot lead to any useful result, especially, but not limited to, the alleged detection of stealth targets.

## 1. Introduction

### 1.1. The International Research Context and the Aim of this Work

The last two decades have seen many theoretical and experimental research activities to apply *quantum technologies* in fields showing potential interest, such as – listed from the oldest to the newest ones – cryptography, computing, communication networks and sensing. The huge related *private* investments are synthesized in Fig. 1 (dashed line) and in Table 1 (data are taken from [1]). Around the 2022 peak, the order of magnitude of public investments - including research funding - is twice that of venture capital, i.e. reaches the notable amount of $ 4-5 billion yearly.

The birth of Quantum Cryptography can be dated back to 1984 with the publication of the celebrated Bennet & Brassard (BB84) method for cryptographic key distribution [2]. But since, in spite of the elapsed thirty-eight years (a time frame which has seen enormous progresses in processing, telecommunications and sensing), practical and industrial applications of Quantum Key Distribution (QKD) and, more generally, of Quantum Cryptography have been much more limited than one could expect two decades ago. As a matter of fact, Quantum Cryptography is not yet widely available to end users, being primarily researched and tested by academic and governmental organizations, and by some industry players, with limited deployments in certain niche applications, for example, by some agencies and financial institutions, with tailored deployments to specific use cases for a small number of users. Moreover, the infrastructure required for quantum communication, including quantum networks and quantum devices, is not yet widely available.

Probably the major effect of research on Quantum Cryptography has been some important investments on the related fields of Quantum Computation and Quantum Networks, see for example *"Quantum Internet Alliance-Phase 1"* (1 October 2022 – 31 March 2026, Total cost 24 M€), [3].

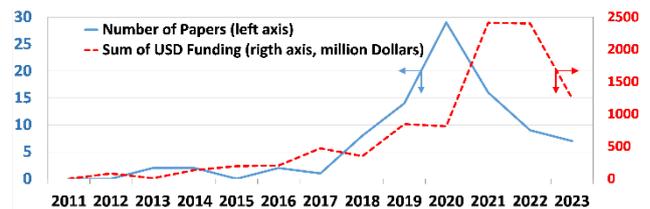

Figure 1. Dashed line - left axis: sum of private USD funding in $ million. (Source: The Quantum Insider, updated end of December 2023, [1]). Continuous line – right axis: papers on Quantum Radar year by year until 07/11/2023 referenced in IEEEXplore. WEB: https://ieeexplore.ieee.org/Xplore/home.jsp. A single book [20] appeared in November 2011.

|      | Americas | EMEA | APAC | Total |
|------|----------|------|------|-------|
| **2022** | 1369 | 762 | 260 | 2391 |
| **2023** | 240 | 781 | 217 | 1238 |

Table 1. Total private investment by region ($ million). EMEA: Europe, Middle East, and Africa; APAC: Asia-Pacific. (Source: The Quantum Insider, updated end of December 2023, [1])

Another relevant application of Quantum Mechanics is found in Metrology [4], with the redefinition of the international system (SI) approved by the General Conference of Weights and Measures on 16 November 2018 and entered into force on 20 May 2019. All seven base units (defined in terms of a fundamental constant of nature) and *volt*, *ohm* and *ampere* were realized by quantum experiments in solid-state devices. This Conference fixed the exact values for the *electron* charge:

$e = 1.602176634 \cdot 10^{-19} A \cdot s$, and for the *Planck*'s constant: $h = 6.62607015 \cdot 10^{-34} J \cdot s$.

For a comprehensive discussion of principle and applications of Quantum Mechanisms, the interested reader may see [5].

Recent years have seen the birth of a sort of a worldwide *"quantum fever"* affecting many domains, arriving to surprising proposals such as heat engines operating at $T \cong 120 \, nK$ [6], and some endoreversible quantum Otto cycle engines [7].

In this *"fever"*, as one could expect, radar researchers (and planners) could not *"sociologically"* remain at the window. In fact, since ten or fifteen years it was, and it is sometimes claimed, that a Quantum Radar (QR) has the potential to outperform classical radar thanks to the properties of quantum mechanics, including in some cases, the detection of *stealth targets* [8], [9], [10], [11] and [12].

Moreover, in [13] we read (comments are left to the reader): *"Billions of dollars have already been spent on quantum computing; compared to this, a few million dollars to develop a field-testable QTMS (Quantum Two-Mode Squeezed) radar does not seem extravagant. With the evidence before us, it seems worthwhile to make a modest effort to understand the possibilities of quantum radars more thoroughly"*.

In [9], we quote: *"Canada has also invested C$ 2.7 m (£ 1.93 m) into developing quantum radar via an ongoing research project at the University of Waterloo"*.

In Italy, the Ministry of Defence – thru his organization named *Teledife* – is financing the research project *"Quantum Radar"*, from 30 April 2022 to 1 May 2025 (https://www.inrim.it/en/research/projects/quantum-radar) [14].

This paper, whose reading does not require a special knowledge of quantum mechanics, is aimed to help the community to better understand the engineering aspects and the intrinsical limitations of the proposed QR technologies and demonstrators as well as some related *"research evaluation"* problems (see *personal communication* in Ref. [15]).

In particular, it is investigated whether a fielded and operating QR system might really outperform an *"equivalent"* classical radar, or not. Here, the term *"classical"* is opposed to *"quantum"* and does include advanced radar architectures such as the Noise Radar (NR) one, possibly the closest to QR. Moreover, it is investigated whether (or not) this QR could reasonably detect a target outside a laboratory, i.e. at least at *hectometer* or *kilometer* ranges, such as the ones of a cheap (order of thousand €) and simple marine radar similar to the one whose characteristics and performance are shown in [16].

**1.2. Literature and Technical Evaluations on Quantum Radar**

The numerical evaluations in this paper consider only the power budget for the Radar Range computations. For the other aspects, although equally relevant (such as resolution, accuracy, resilience to the jamming, …) the interested reader is addressed to [17], [18] and to the numerous References of the latter.

The characteristics of certain proposed types (see for instance [13] and [19]) of Quantum Radar (sometimes the term *"quantum protocol"* is used in place of *"type"*) make them allegedly similar to a Noise Radar (NR) [17], and a clarification is done in this frame, too.

A search (done on November 7[th], 2023) for *"quantum radar"* on the well-known IEEE database IEEEXplore, which includes more than 6.1 million items, yielded the results shown in Fig. 1 (continuous line). Overall, one finds about 90 papers (on Journals, Reviews or Conference Proceedings) and one book, [20]. Including the papers not referenced in IEEEXplore, the total exceeds one hundred publications in 2012 – 2023. Fig. 1 shows a decreasing number of papers on QR from 2020 to 2023.

Most QR papers are oriented to quantum physics and to technological aspects, with a very few contributions (order of 4 or 5 %) considering system (and of course, operational) concepts. Operating demonstrators are practically absent: sometimes, in spite of the word *radar*, some described laboratory tests refer to optical wavelengths (i.e. to a Lidar, not to a Radar); anyway, most tests are in the lab, not outdoor.

In one case the *"quantum radar demonstrator"* – as shown in Fig. 3 of [19] – is made up by a transmitter connected to a horn antenna facing another (receiving) horn antenna at a distance of less than one meter (both horns being fixed with adhesive tape on a desk, neglecting multipath and reflection effects), i.e. with no radar target at all. Hence, the experiment shown in [19] was carried out with a one-way propagation attenuation $a_R$ (see below) not measured but probably close to the unity.

Note that, in a real radar operation (e.g. for plane tracking applications), the two-way attenuation is of the typical order of $10^{-13}$ as a ten watts transmitted signal generates echoes of the typical order of picowatt or less at normal (kilometric) target distances. Conversely, in the aforementioned literature one founds a much lower attenuation: see for instance, among the recent papers, [21] where the results shown in Figs. 2 and 3, according to their captions, are obtained with a round-trip transmissivity of mere two orders of magnitude, i.e. $0.01 \, (-20 \, dB)$. This attenuation corresponds to a radar Range of the order of one metre (in the X-band, using horn antennas and for a target of $1 \, m^2$ radar cross section).

**1.3. Outline of this work**

Detection and Ranging capabilities for QR are critically discussed and, after an analysis of the global research situation in this area, a comparison with its *closest* Classical Radar, i.e. the Noise Radar, is presented.

Finally, conclusions are drawn on the potential advantages of a QR versus its Noise Radar counterpart and, more generally, on the QR research.

Some widely known papers on QR lead us to emphasize the importance of the Order of Magnitude (OoM) and, *last but not least*, of the Common Sense.

As a matter of fact, from the aforementioned literature, these canonical concepts seem not to be always obvious.

## 2. Historical Premises and Present Situation of Quantum Radar

### 2.1. A Short History of Quantum Radar

Despite its appearance, the idea of *quantum radar* is not new (the history is resumed in [22], [23] and [24]) as the concept – with a description of a possible embodiment – can be found in an old (2008) USA patent (assignee: Lockheed Martin Corporation) [25]. Incidentally, it is worth remarking that in the last fifteen years neither Lockheed Martin nor any of the big Radar and Defence Companies, to the best of our knowledge, have developed or anyhow exploited this invention, a situation, anyway, which applies to about 50% of the patents worldwide.

The patent [25], in its paragraph 5.1.3, describes three classes of quantum sensors. The third one *"transmits quantum signal states of light that are entangled with quantum ancilla states of light that are kept at the transmitter"* and Fig. 5.2 of the patent shows *"an entangled pair of photons, one stored in the transmitter and another reflected from the target, sent to a measuring device"*.

Most of research on Quantum Radar is based upon the above concepts, within the obvious, universal understanding (although not well clarified in the above patent) that the *"reflected"* photon, having interacted (at least) with a complex object, i.e. with the target, loses the entanglement, and that amplification (in the transmission or reception paths) destroys the entanglement.

Moreover, some related literature (see [26], [27], [28]) shows that the radar cross section of a target (apart the back-scattering side lobes) shall not significantly change passing from the Classical Radar (CR) to the QR, in similar operating conditions (in reality it does not change at all, as clearly shown in [29]).

The readers interested to get an overview of the research on QR as at 2020 may refer to [30]. An overview on the basic concepts in Quantum Mechanics is presented in Chapter 2 of the book by V. J. Stenger, [31], a book full of nice, reasoned criticism.

### 2.2. Quantum Radar Today

Two main classes of quantum radars (neglecting the outdated *interferometric quantum radar* referenced in [20]) have been proposed in the open literature: *quantum illumination* (QI) radar and *quantum two-mode squeezing* (QTMS) radar [19].

The concept of *"quantum illumination"* (with the related protocol) was introduced in 2008 [32]; it starts with a generation of pairs of *entangled* photons, the *idler* and the *signal* photon. The signal photon is sent to region where a target could be present, while the idler is stored. If a target is present, the signal photon may be received by the radar after the transmission delay, otherwise the radar only receives noise photons. Each received photon is compared with the idler in some kind of measurement. Of course, in addition to the basic detection function, a radar set is (at least) requested to measure the distance of the targets, i.e. the ranging, which is not trivial using the quantum illumination protocol, as quoted in [33]: *"... hinging on a joint-measurement between a returning signal and its retained idler, an unknown return time makes a Quantum Illumination-based protocol difficult to realise"*.

Aimed to solve this problem, the QTMS protocol, which operates in a way closer to that of a conventional radar, has been implemented as a laboratory demonstrator (but with no target), [13], [19], [24]. It circumvents the ranging problem as follows. In the QTMS radar, the reference-entangled beam is immediately measured using heterodyne $I$ (in-phase) and $Q$ (quadrature) detection and retained within the system, while the received signal is measured at its arrival. Hence, the reference and the received signals are used to simultaneously measure their correlation.

According to quantum theory, the QI protocol should yield better results, but storing the reference signal until the arrival of the corresponding echo signal is really difficult, especially at radar (microwave) frequencies. Hence, the QTMS radar is mainly considered here. It requires maximally entangled pairs of photon modes; therefore, the process of spontaneous parametric down-conversion is the most generally used, generating a Gaussian two-mode squeezed-vacuum state at microwave frequencies.

Despite the loss of the entanglement due to the interaction with the environment (including the amplification), QTMS radar is aimed to exploit the correlation caused by the entanglement to detect the signal photons in noise when the correlation is computed many times. The number of pairs (or *"of pulses"*) $M$, i.e. the number of distinguishable modes, is referred to as the time-bandwidth product: $M = B \cdot T$, where $T$ is the duration of the emitted signal (less than, or equal to, the time-on-target) and $B$ is its bandwidth. In each mode, an average number $N_s$ of photons is transmitted; as for $N_s \gg 1$ the classical physics applies, a quantum advantage is fully attained when $N_s \ll 1$.

Among the (not numerous) experimental evaluations, in 2020 Barzanjeh et. al. [34] carried out experimental verification of Quantum Illumination in X band, with generation and amplification of entangled microwave photons (frequencies: $10.09 \, GHz$ and $6.8 \, GHz$) in cryogenic conditions (at $7 \, mK$) and with a target at room temperature and at a fixed distance of one metre. The experiments showed $1 \, dB$ advantage over the optimal classical illumination at $N_s < 0.4$, the difference with respect to the theoretical $3 \, dB$ being explained by the limitations due to the experimental set up.

Some recent technological developments, related to QR, include the wide-band Josephson Traveling Wave Parametric Amplifier [11], [12] and the optical technologies, [35], [36]. A proposed general scheme of optical quantum radar [37] uses, in transmission, an electro-optical down-conversion, permitting to create the entangled photons in the optical region, and to down convert them to microwave photons; in reception, an up-conversion back to the optical region for detection.

Today the basic *"Quantum"* part of a QR set is the generator of microwave-entangled photons. A typical implementation is described, inter alia, in [38] where, at Section II, one finds the description of the operation of a Josephson Parametric Amplifier (a microwave resonant cavity terminated by a Superconducting Quantum Interference Device, or SQUID) and of the need to keep it very close to the *absolute zero temperature* (i.e. at a few *milli-Kelvins*) within a bulking dilution refrigerator.

The latter is described in [38] as having the size of a large car including the He-3 and He-4 large Dewar's, and his power consumption is alleged as large as $15\ kW$; its high cost (order of $10^6$€ according to [39]) causes a of the radar cost of the QR set (Fig. 1 of [39]) to be *five orders of magnitudes* greater than the equivalent conventional radar.

In front of the significant *SWaP* (Size, Weight and Power) implicit in the QR technology, one may ask what radar performance enhancement arise from the Quantum approach. In the literature there are many theoretical evaluations, indicating a gain (depending on the quantum protocol and on the average number of signal photons $N_S$ and of noise (background) photons $N_B$, see Section 3) up to $3\ dB$ or $4\ dB$ or $6\ dB$ (the highest figure is the one generally cited).

An overview of the attained Quantum Advantage as at May, 2020 is found in [40] with a synthesis of experiments in Table 1 of [40]. Note that in a conventional monostatic radar the link budget may be improved by $6\ dB$ increasing the antenna size 1.41 times or the transmitted power four times, with a much more limited SWaP increase then going to QR approach.

Moreover, the evaluations in Section 3 show that the *"long-distance detection"* of QR in the title of [12] will never apply to real world situations, in spite of the known or anticipated technological improvements.

Of course, similar results apply to the alleged detection of *stealth targets*, see [9] and [11]. In the Abstract of [11] we read: *"... making our MQI (Microwave Quantum Illumination) system a promising candidate for the detection of stealth objects"*. In reality, stealth targets call for a high power illumination which is contrasting with the QR nature itself, as explained later, and the radar cross section of a target (either stealthy or not) does not change when QR is used, as elegantly shown in [29], as the path of each photon to the target is not well defined because of the position uncertainty, and this causes some quantum interference which exactly replicates, in the far-field limit where the radar cross section is defined, the classical scattering behavior of electromagnetic waves.

### 2.3. Quantum Two-Mode Squeezing Radar and Noise Radar

In [13], [19], [24], [41] it is claimed that the above-described QTMS radar operation is similar to the one of a Noise Radar (NR) [42], [43]. In reality, NR and QR are quite different as one may see comparing Fig. 2 e Fig. 3 for the NR with Fig. 2 of [19] for QTMS radar. In section 4 a short history of the NR will be proposed together a comparison with QR concerning the maxim range.

In [19] a comparison is done with a particular (and quite *"artificial"*) classical radar demonstrator, named Two-Mode Noise (TMN) radar and implemented as close as possible to the QTMS radar demonstrator (including the cryostat refrigerator at a few $mK$), but with the pair of signals generated by mixing Gaussian noise with a sinusoidal carrier, i.e. not being entangled.

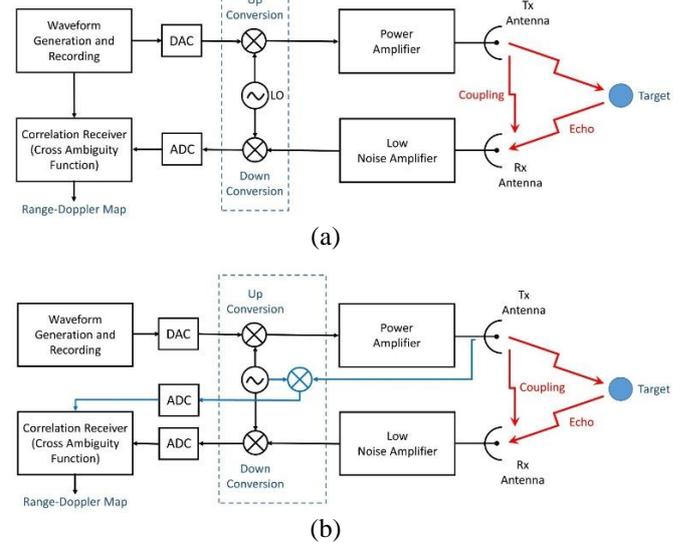

(a)

(b)

Figure 2. Basic Block Diagram of a Noise Radar. (a) The reference is the digital record of the transmitted code. (b) The reference is the record of the transmitted signal at the antenna port. ADC = Analog-to-Digital-Converter. DAC = Digital-to-Analog-Converter.

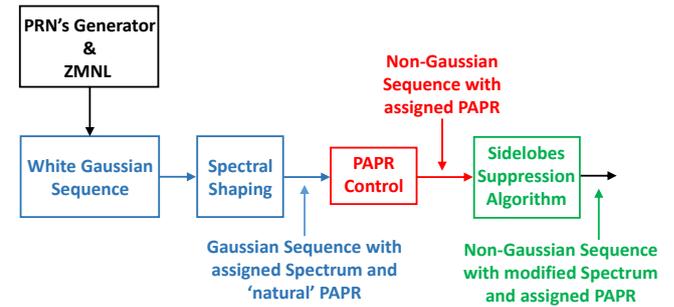

Figure 3. Basic Block Diagram of the waveform generator for a modern Noise Radar. PRN: Pseudo Random Number, ZMNL: Zero Memory Non Linearity. PAPR: Peak-to-Average Power Ratio.

In some experiments, when both systems transmit in the air signals at $-82\ dBm$ (from a horn antenna to a close-by similar antenna) it resulted that the QTMS radar required one eighth of the integrated samples of the TMN radar to achieve the same performance in terms of Receiver Operating Characteristics (ROC curves).

However, the Authors of [19] warn that a similarity between a NR and the TMN radar demonstrator stays only in the fact that both NR and TMN transmit random signals. In this frame, it is very important to remark that the randomness in the TMN radar (and, more generally, in QR) is unavoidable and uncontrollable, being due to the quantum-mechanical signal generating process. In modern NR the preferred solution is full digital and uses pseudo-random number (PRN) generators [42] - [46], see the high-level block diagrams of Fig. 2 and in Fig. 3 and the following comparison among a Conventional Radar (a Noise Radar) and Quantum Radar.

The consequences of these different generation processes of QR and NR include a poorer detection performance of the QR, as discussed in the ensuing Sections.

## 3. Radar Range for Quantum Radar

### 3.1. General Remarks on Bose-Einstein Statistics and Noise Background

Determining the maximum operational Range of a radar set seems feasible with a simple computation of the Radar Equation, historically standardized by the classical report (and ensuing book) on Pulse Radar Range by L. V. Blake, [47]. However, the task is more difficult when considering the so many factors affecting the computation (target fluctuations, equipment losses, multipath, internal and external disturbances and more), so much that the real operational radar Range sometimes is as short as half of the computed Range, i.e. detection losses whose sum is as large as $12\ dB$ are neglected. This is probably a worst-case (mentioned by M.I. Skolnik, [48]) but errors as large as $3 - 4\ dB$ are probably very common.

The maximum Range of a QR is an even more complex matter being related to the statistical description of the involved photons. From the Bose-Einstein statistics (as applicable in the cases of *zero chemical potential*), the average number of photons per mode $N_B$ (where the subscript $B$ stands for background) versus the frequency $f$ at a system temperature $T_s$ (in Kelvin) is:

$$N_B = \frac{1}{exp\left(\frac{hf}{K_B T_s}\right) - 1} \quad (1)$$

with $K_B = 1.38065 \cdot 10^{-23}\ J/K$ the Boltzmann's constant and $h = 6.62607 \cdot 10^{-34}\ J \cdot s$ the Planck's constant. At radio and microwave frequencies and at room temperature $hf \ll K_B T_s$, hence, considering the approximation of the exponential at the first order: $exp\left(\frac{hf}{K_B T_s}\right) \cong 1 + \frac{hf}{K_B T_s}$, Eq. (1) becomes: $N_B \cong \frac{K_B T_s}{hf}$, from which the background noise power inside the bandwidth $B$, is equal to the classical relationship:

$$P_{rn} = K_B T_s B \cong N_B hf B \quad (2)$$

Fig. 4 shows Eq. (1) varying $f$ from $0.1\ GHz$ to $100\ THz$. The blue area is the optimal for Quantum operation; the QR system has the optimum Quantum Advantage for an average photon number $\ll 1$ and loses the advantage for a number of photons greater than about five (grey area). Dashed lines represent the linear (*classical*) approximation $N_B \cong \frac{K_B T_s}{hf}$.

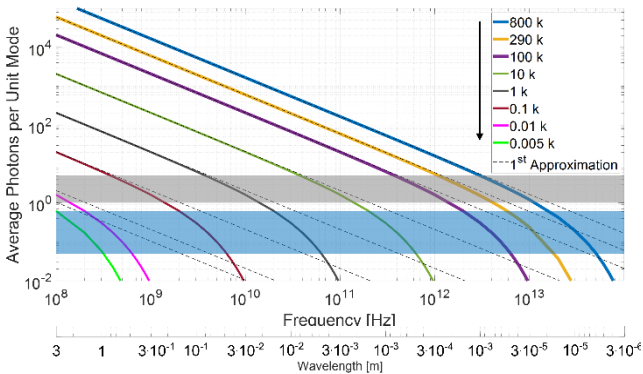

Figure 4. Average photons per unit mode, Eq. (1), versus the frequency (up to infrared radiation, $100\ THz$). $T_s = 0.005, 0.01, 0.1, 1.0, 10, 100, 290, 400, 800\ K$. Dashed lines show the first order approximation.

For a thermal source at $3000\ K$, the number of photons per mode is less than unity, and of the order of $3 \cdot 10^{-4}$ in the middle of the visible spectrum ($5.4 \cdot 10^{14}\ Hz$, or $\lambda = 0.55\ \mu m$). For an X band radar at room temperature $N_B$ is of the order of $10^4$, and is less than the unit only at $T_s$ well below $1\ K$.

### 3.2. Range computations for proposed Quantum Radars - Comments

The literature on the Range performance of a QR, often, simplifies the disturbance as *"background noise"* with $N_B$ photons, while radar engineers know very well that the radar disturbance is something more complex. It includes unwanted echoes, propagation effects (with related attenuation), antenna noise, radiofrequency connections to the receiver, and finally, the first active reception stages, [47], [48] and [49].

In [50] we read (QI stands for Quantum Illumination, $\hbar = \frac{h}{2\pi}$, $W$ is the bandwidth): *"In QI radar, the noise power of the receiver can be expressed as $P_{rn} = \frac{\hbar \omega W}{exp(\hbar \omega / K_B T) - 1}$, where $K_B$ is the Boltzmann constant and $T\ (= T_s)$ is the equivalent noise temperature in Kelvin. In practice, it is difficult to accurately calculate the noise temperature, which is related to factors such as antenna geometry, beam direction, solar activity, and signal wavelength etc. Therefore, in the subsequent analysis, we define the equivalent noise temperature of $3 - 300\ K$, which can represent the majority of radar operating scenarios"*.

The lower value is quite optimistic: the system noise temperature $T_s$ of a radar set is typically close to, or above, $250 - 300\ K$. In fact, the system noise temperature $T_s$, [49], referenced to the output of the radar antenna, is the sum of three contributions: by the antenna, by the radiofrequency (RF) connections of the antenna with the receiver (including the duplexer - a transmit-receive switch needed in monostatic radars - and the rotary joint - if any) and, finally, by the receiver itself (whose main element is normally a Low Noise Amplifier, LNA):

$$T_s = T_a + T_{RF} + T_{LNA} \quad (3)$$

with:

- $T_a$ resulting from the external noise sources including the sun, the cosmic background at $2.7\ K$, the atmosphere and the land and sea surfaces. It is a highly variable quantity; common graphs [47] supply this contribution versus the operating frequency for a *"standard environment"* and for different values of the pointing angle $\theta$ of the antenna with respect to the vertical. For example, in the X band (around $9\ GHz$) $T_a$ has a maximum value of about $100\ K$ when pointing towards the horizon ($\theta = 90$ degrees) and a minimum value of about $10\ K$ in the unrealistic case of zenith pointing ($\theta = 0$). To set exemplary values, assuming $\theta = 30$ degrees, we have $T_a = 30\ K$ (but for a Surface Movements Radar (SMR) whose antenna points down, $T_a$ is much higher).

- $T_{RF} = (L_{RF} - 1) \cdot T_0$, being $T_0$ the reference temperature of $290\ K$ (according to the IEEE standard)

or, if known, the physical temperature of the previously mentioned RF connections, and $L_{RF}$ their attenuation (i.e. the loss). An exemplary value (for a $0.5\ dB$ loss) is $T_{RF} = 35\ K$.

- $T_{LNA} = (F - 1) \cdot L_{RF} \cdot T_0$, where $F$ is the Noise Figure of the amplifier. For an exemplary $F = 1$ dB and the above 0.5 dB loss, $T_{LNA} = 190\ K$.

The sum yields $T_s = 255\ K$ (but over twice for a $SMR$).

There are various QR approaches, or *"protocols"*: for a general discussion, see [18], and a general overview of the Quantum Radar principle from its inception can be found in [22]. In the latter, it is explained that: *"… despite loss (rectius: losses) and noise that destroy its initial entanglement, quantum illumination does offer a target-detection performance improvement over a classical radar of the same transmitted energy"*.

As shown in Section 2, a Quantum Radar transmits a sequence of $M$ modes i.e. high time–bandwidth product single-photon signal pulses, each of which is entangled with a companion single-photon idler (or reference) pulse. The receiver makes a decision between the hypothesis of target absent and the one of target present.

Basic and contrasting facts dominate the power budget, hence the computation of QR Range. They are:

a) The energy in a single photon at microwave or millimetre-wave frequencies is extremely small when compared to the one of a Conventional Radar (CR) pulse: therefore, being $M$ defined by operational constraints, one could try to increase the number $N_s$ of signal photons per mode. This increase brings back the radar system to the classical operation. With a number of photons per mode, say 0.01, optimal for quantum-advantage, the transmitted energy per microwave radar pulse (i.e. per mode) is of the order of 0.01 femto Watts, i.e. 16 to 20 orders of magnitude below what is required for target detection. Moreover, according to quantum mechanics, the amplification of the radar signal generates noise, which would nullify the quantum advantage, [40].

b) Theoretically, a Quantum Illumination (QI) system shall provide a factor-of-four ($6\ dB$) improvement in the error-probability exponent over its classical counterpart of the same transmitted energy [40], [41]. However, this improvement can be obtained, remaining in the Conventional Radar (CR) technology (hence avoiding cryogenic generators) increasing the dimensions of the Transmit/Receive antenna (antennas), for example, in the monostatic case (a single Tx/Rx antenna), changing a $1.2\ m$ dish into a $1.7\ m$ dish. The practical QR implementations limit this advantage to lower figures, order of 1 to 3 $dB$ only, [40], [51].

c) The benefit of QR over CR is significant for a very few (less than the unity) photons per mode and disappears for more than a few transmitted photons per mode. From theory and experiments, it results that there is a negligible benefit of a QR with more than five photons per mode.

d) The energy of a photon is proportional to its frequency, calling for QR operating in the millimetre-wave or terahertz bands, where, unfortunately, the atmospheric attenuation prevents long-Range operation.

e) The increase of the number of modes $M$ for the (necessarily limited) available signal bandwidth $B$ would generate an increase of the pulse duration, i.e. of the dwell time $T$. Values of $T$ above some threshold (order of a few milliseconds to hundreds of milliseconds depending upon the type and dynamics of the particular target) would render the system prone to the effects of target scintillation and of Doppler frequency, destroying the correlation with the stored replica, hence the quantum advantage.

Summing up, we quote from the Introduction of [51]: *"Our main conclusion is that, while realizable experimentally, useful application of microwave quantum radar protocols to any conventional setting is unrealistic because of fundamental restrictions on power levels"*.

Recently the use of QR has been proposed for biomedical sensing [38], i.e. a short range (order of meters or less) case. Short-range radars were introduced for healthcare applications of detecting human vital signs in 1975. Heart rate measurements have been successfully measured at $1\ m$ using sub-micro-watt power levels, [52]. Non-invasive microwave techniques for contact and remote sensing of respiratory and circulatory movements have been developed at continuous-wave frequencies between 1 and 35 $GHz$ with the average power density of energy radiated ranging from $1\ \mu W$ to $1\ mW$ per *square cm*, i.e. much lower than the ones due to the cellular phones used by patients and sanitary personnel. Some systems are capable of measuring pressure pulse, heart rate, and respiration rate in contact with body surface or at distances greater than $30\ m$, or behind thick layers of non-conductive walls.

In biomedical and healthcare applications, the QR is not an option (contrary to what is written in [38]), because:

- SWaP limitations are important.
- The short distances imply very low transmitted microwave power levels for which a 6 dB advantage is immaterial.

### 3.3. Exemplary Range computations for Quantum Radar

Some computations for a representative QR are shown in the following to sustain the previous discussion.

The related main parameters are:

- $f_0$: operating (central) frequency;
- $\lambda = c/f_0$: wavelength;
- $B$: operation bandwidth, i.e. radar frequencies from $f_0 - \frac{B}{2}$ to $f_0 + \frac{B}{2}$;
- $T$: signal duration (less or equal to the dwell-time);
- $M = B \cdot T$: number of modes;
- $\sigma$: radar cross section of the target;
- $G$: antenna gain (the same for Tx and Rx antenna);
- $T_s$: system noise temperature;
- $SNR$: signal-to-noise ratio;

- $a_R$: free-space attenuation of the radar equation at distance $R$ (the same in both ways).

First, the free-space attenuation is:

$$a_R = \frac{G^2 \lambda^2}{(4\pi)^3 R^4} \sigma \quad (4)$$

Hence, the received power from a target at a distance $R$ is:

$$P_{rs} = M \cdot N_s h f_0 B \cdot \eta_Q \cdot a_R \quad (5)$$

where $N_s$ is the average number of photons per mode and $\eta_Q$ the quantum advantage alleged to range from 0 to 6 $dB$ [34]. Hence, using Eq. (2) and Eq. (5), the signal-to-noise ratio is:

$$SNR = \frac{P_{rs}}{P_{rn}} = M \frac{N_s \cdot \eta_Q}{N_B} \cdot a_R = B \cdot T \frac{N_s \cdot \eta_Q}{N_B} \cdot a_R \quad (6)$$

Hence, to achieve a *positive* (in $dB$) $SNR$ at Range $R$, the Quantum Radar shall operate with a time duration:

$$T \geq \frac{N_B}{N_s \cdot \eta_Q} \cdot \frac{1}{B \cdot a_R} \quad (7)$$

At the widely used X band ($f$ order of $9 - 10\ GHz$) e.g. at frequency $f_0 = 9.37\ GHz$ ($\lambda = 0.032\ m$), assuming a bandwidth $B = 1\ GHz$ (that is about 10 % of the centre frequency $f_0$), for the sake of simplicity we set $N_s \cdot \eta_Q = 1$, $\sigma = 1\ m^2$, at $R = 1000\ m$ with an antenna gain (the same in transmission and in reception) of $1000$ ($30\ dB$) and with $T_s = 400\ K$, it results: $a_R \cong 5.16 \cdot 10^{-13}$ (which is about $-123\ dB/m$) and $N_B \cong 889$. Hence, Eq. (7) gives: $T \geq 479\ h$ (order of days!) which appears absurd. For $T_s = 290\ K$ and $T_s = 255\ K$, it results $T \geq 367\ h$ and $T \geq 323\ h$ respectively.

We underline that the use of electromagnetic spectrum by radar is regulated by the ITU (an ONU agency) and the radar bandwidth allocation generally does not exceed 10 %. Therefore, the theoretical possibility of radar transmissions in an ultra-wide band is limited to indoor, very short-Range, applications [53].

In the ideal case of $1\ m^2$ target much closer, i.e. at $R = 10\ m$, the $SNR$ is $10^8$ times greater than at $1\ km$ and an operation with same $BT$ should permit $T$ to be at the order of magnitude of ten milliseconds.

These approximate results agree with those in Fig. 1 of [51] (to be scaled down from $BT = 10^9$ to $BT = 10^6$ or $10^7$) and in [54], confirming that in the microwave region (or below) the maximum range of a QR shall not exceed the order of a few metres. From Eq. (6) and Eq. (4) the maximum range ($R_{max}$) can be evaluated as:

$$R_{max} = \left[ \frac{G^2 \lambda^2 L \cdot \sigma}{(4\pi)^3 SNR_{min}} \cdot \frac{N_s \cdot \eta_Q}{N_B} B \cdot T \right]^{\frac{1}{4}} \quad (8)$$

where $\eta_Q$ is the quantum advantage and $L$ the total loss, neglecting the attenuation of the medium. Setting: $L = -4\ dB$, $SNR_{min} = 13.2\ dB$, $\sigma = 1\ m^2$, $N_s \cdot \eta_Q = 1$, $\lambda = 0.032\ m$, $B = 1\ GHz$, varying the time duration $T$ from $1\ ms$ to $1\ s$ (i.e. $M$ from $10^6$ to $10^9$), Fig. 5 shows a comparison between the maximum ranges at the same operating conditions. For the Quantum radar the maximum range is the order of meters, while Noise Radar improves the Range (see also Eq. (9) in section 4.3) to the order of ten kilometres (the system temperature is set to $100, 290, 800\ K$). For the Quantum radar, Ranges greater than about ten or twenty meters can only be achieved when the whole radar set, including the antenna (and the external surfaces within its main lobe) is cooled at cryogenic temperatures, which is not compatible with any use outside a specific laboratory.

Note that the Quantum Radar situation improves when the frequency increases ($N_B$ quickly decreases, see Fig. 4), but above circa $8\ GHz$ the attenuation by rain becomes a critical factor and, with increasing frequency above $35\ GHz$, the attenuation by the atmosphere also becomes critical, preventing QR for any operational *"outdoor"* either civilian or military radar application.

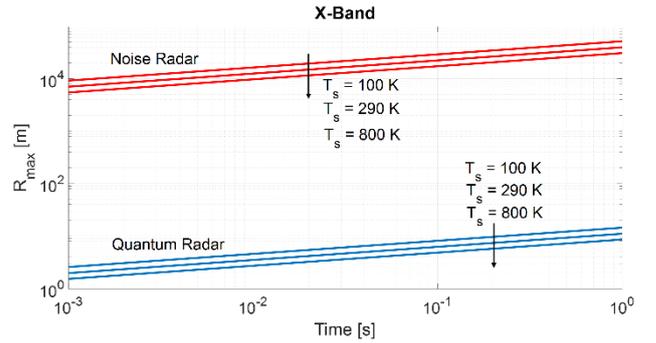

Figure 5. Comparison between the Maximum range for Noise Radar, Eq. (9), and Quantum Radar, Eq. (8), at X-band vs time-duration $T$. The Noise Radar transmitted power is 100 mW (for 10 μW the lines are shifted below by a decade), $G = 30\ dB, \sigma = 1\ m^2, L = -4\ dB, B = 1\ GHz, SNR_{min} = 13.2\ dB, N_s \cdot \eta_Q = 1$. No atmospheric attenuation, no clutter, no radiofrequency interference.

Evaluations of detection performance at $35\ GHz$ (Ka band) and $95\ GHz$ (W band), not shown here, conform an order of magnitude of the maximum Range for QR below *ten* meters with a dwell time below one tenth of second, similar to the results shown in Figs. 2 and 3 of [50].

## 4. Quantum Radar vs Noise Radar

### 4.1. Some history of Noise Radar

The history of Noise Radar (NR) is quite old (for the related References, please see [44]). Noise radar was introduced in 1959 by Horton for a high-resolution distance measurement system; the generation of noise signals was first implemented using analog sources; for instance, generation of "chaotic" signals from an analog source at W-band was developed in Ukraine in the '90s.

Other research was performed in China, in the USA, and in Europe. Since the 2000s, experimental research activity with field trials took place in Ukraine, where in 2002, the First International Workshop on Noise Radar Technology (NRTW 2002) was held.

In Poland, at the Warsaw University of Technology starting from 2010's, a noise radar demonstrator was implemented using commercial hardware for the detection of moving targets.

From 2005, a noticeable pre-competitive and unclassified research effort on NR is being developed in the frame of the NATO Sensors and Electronic Systems (SET) Research Task Groups, RTG's. From December 2020 the activities continue within the RTG SET-287 "Characterization of Noise Radar".

**4.2. Noise Radar: Peak-to-Average Power Ratio and Range Sidelobes Level**

In some works (as said in section 2, see for instance [19] and [24]) QR has been associated to NR, due to the common feature of randomness of the transmitted signals.

However, there is a basic difference in the generation of the signal. In QR the signal is *"naturally"* random and not modifiable, while in the present digital version of NR, the signal can be *"tailored"* [43], [44], see Fig. 3.

Since long time it is well known that radar detection depends on the energy from the target's echo rather than on its power, as the output of the optimum (or *"matched"* or *"pulse compression"*) filter [55] is proportional to the energy of the received waveform divided by the spectral density of the noise ($E/N_0$). Hence, most *classical* radar waveforms have a constant amplitude (i.e. are phase-coded or simply not coded at all) in order to exploit at best the power amplifier, granting a Peak-to-Average Power Ratio ($PAPR$, [56], [57], [58]) equal to the unit to maximize, with the bound of the maximum transmittable power, the received energy in the dwell time.

Note that *classical* pulse-compression radars may transmit either deterministic or random/pseudorandom waveforms. In the latter case, the randomness of the transmitted waveform, with its $PAPR < 1$, may significantly reduce its energy, affecting the detection performance [56].

Radar operation with time-bandwidth product $BT > 1$ poses the problem of Range-sidelobes at the output of the coherent integration (i.e. of the compression filter). In the Noise Radar case, the problem may be faced by a suited *"tailoring"* of these signals, [44], allowing a Peak Sidelobe Level ($PSL$) as low as (typically) $-60\ dB$ below the main lobe and a $PAPR$ suited to the power budget, e.g. as low as 1.5 (corresponding to a $1.76\ dB$ loss).

Conventional Radars (CR) transmit a pulse (see Fig. 6a). The correlation is approximated by a narrow band filter with a bandwidth close to the inverse of the pulse width. In the case of pulse compression, a coded waveform is used (MOP: Modulation on Pulse in the Electronic Defence jargon) and the received signal is (nowadays, digitally and often in the Fourier domain) correlated with a template of the transmitted one.

In Noise Radar (see Fig. 6b), the received signal is correlated with a template of the transmitted one, which is created by a (possibly, tailored) realization of a random process, in turn, obtained by a noise source or, more frequently, by a pseudorandom numbers generator.

In Quantum radar (see Fig. 6c), the received signal is correlated with the idle signal entangled with of the transmitted one, which is a (not-controlled) realization of a Gaussian random process.

By comparison, of cases in Fig. 6b and Fig. 6c, it is clear that the main differences are:

i. Noise Radars may transmit every kind of pseudorandom signals, including those with constant modulus (phase-only code) and more generally with non-Gaussian statistics.
ii. In Noise Radars the decision on target, present/absent is made on the basis of the output of the correlator (matched filtering) while in Quantum Radar it is preferably made on the basis of the correlation coefficient.

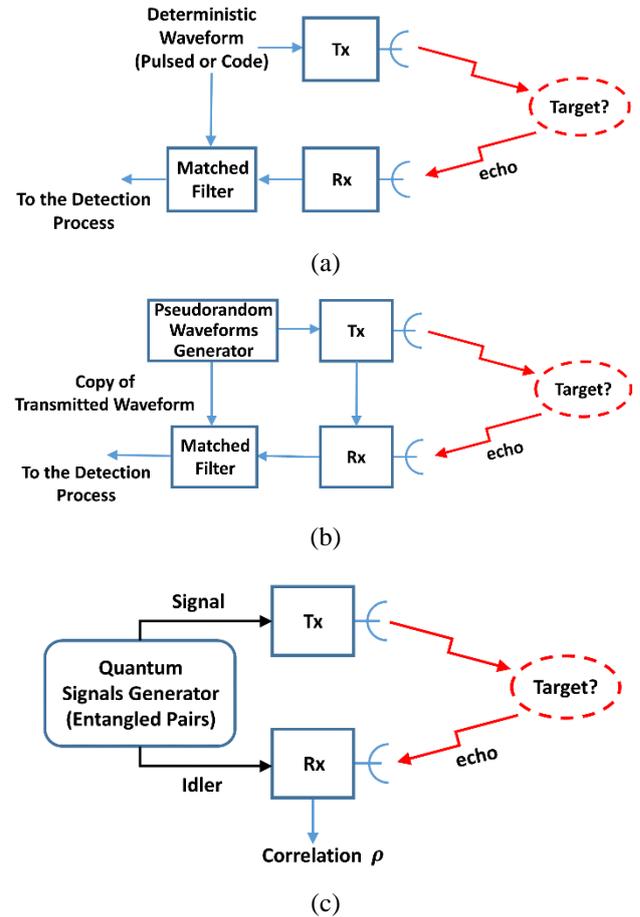

Figure 6. Block-diagram comparison of: (a) Conventional Radar, (b) Noise Radar and (c) Quantum Radar.

**4.3. Maximum Range: comparison Quantum Radar - Noise Radar**

In some papers, [59], [60], Quantum Radar is proposed because of its Low Probability of Intercept (LPI) features due to the intrinsic randomness of its emission. Similar characteristics belong to Noise Radar [42], [43], [44], [45] and [46]. Hence, it is interesting to compare the performance of NR and QR in equivalent system configurations.

A simple comparison figure is the ratio of maximum Ranges: $\frac{R_{NR}}{R_{QR}}$. For Quantum Radar $R_{QR}$ is computed by Eq. (8), while for the Continuous-Emission Noise Radar [44], we have:

$$R_{NR} = \left[ \frac{P_T G^2 \lambda^2 L \cdot G_{INT} \cdot \sigma}{(4\pi)^3 SNR_{min} K_B T_S B} \right]^{\frac{1}{4}} \qquad (9)$$

with the usual meaning of symbols, where $G_{INT}$ is the coherent integration gain, equal to $B \cdot T_{INT}$, with $T_{INT}$ the coherent integration time ($T$ in QR), equal or less than the dwell-time. Hence, the desired ratio is:

$$\frac{R_{NR}}{R_{QR}} = \left[\frac{P_T}{K_B T_s B} \cdot \frac{N_B}{N_s \cdot \eta_Q}\right]^{\frac{1}{4}} \qquad (10)$$

Assuming $N_s \cdot \eta_Q \cong 1$ (in fact, $N_s < 1$ and $\eta_Q > 1$), and taking into account that the antenna and the receiving parts operate close to a room temperature, hence $T_s \gg 1\,K$ ($N_B \cong \frac{K_B T_s}{hf}$, see also Fig. 4) Eq. (10) becomes:

$$\frac{R_{NR}}{R_{QR}} = \left[\frac{P_T}{hf} \cdot \frac{1}{B}\right]^{\frac{1}{4}} = \left[\frac{E_T}{hf} \cdot \frac{1}{M}\right]^{\frac{1}{4}} = \left[\frac{N_T}{M}\right]^{\frac{1}{4}} \qquad (11)$$

where $E_T (= P_T T)$ is the energy coherently transmitted on the target, $M = B \cdot T$ is the number of modes and $N_T$ is the related number of photons transmitted by the NR: $N_T = \frac{P_T \cdot T}{hf}$.

For $P_T = 100\,mW$, $f = 10\,GHz$ it results: $N_T = 1.51 \cdot 10^{22} \cdot T$. To get $R_{NR} = R_{QR}$ one has to set $M = N_T$, i.e. an unthinkable bandwidth $B = 1.51 \cdot 10^{13}\,GHz$, that could only be achieved operating at unrealistic carrier frequencies above $10^{23}\,Hz$.

A similar evaluation is presented in [59] where, however, cooling of both CR and QR sets at $10\,mK$ is considered and the ratio between Eq. (12) and Eq. (14) of [59] leads to a Conventional Radar/Quantum Radar Range ratio equal to: $\frac{R_{CR}}{R_{QR}} = \left[\frac{1}{M}\right]^{\frac{1}{4}}$ (note that in [59] the number of modes is $m$ in place of $M$) which is not in agreement with the evaluations shown in this work, likely, there are errors in the computations (called *"simulations"*) of [59], invalidating its conclusions and the sentence in its Abstract: *"It is shown that the detection range of quantum illumination radar is larger than that of classical radar…"*.

Likely, the pulse compression gain was not taken into account, but it is necessary to mention the recent paper of the same Authors [60], where the compression gain of CR is considered and it is confirmed that the conditions $N_s \ll 1$, $N_B \gg 1$ and $M \gg 1$ maximize the advantage of Quantum Illumination but unavoidably lead to very short radar Ranges.

The conclusions of [60] include: *"… although QI shows its advantages, this advantage is limited to the case of very weak transmitted signal power, so it may be a challenge for applying QI to radar remote detection"*.

In fact, from Eq. (11) one easily computes the frequency $f^*$ making $R_{CR} = R_{QR}$. Posing $B = a \cdot f$, with $a < 1$ (e.g. $a = 0.1$), it results: $f^* = \sqrt{\frac{P_T}{h \cdot a}}$, confirming that Quantum Sensing tends to became useful at very high frequencies (i.e. in the extreme UV or X-rays regime and above) and at very low power levels.

## 4.4. Specific Noise Radar's Advantage over Quantum Radar

Differently from Classical Radar and NR, QR signals cannot be *"tailored"* and are inherently random with Gaussian distribution, thus causing a relatively large peak sidelobes ratio after pulse compression, order of $1/M$.

Important for the radar Range point of view, QR signals have a poor *PAPR*, whose estimated value depend on the chosen truncation point for the Gaussian law, and is of the order of ten or twelve. The related loss, around ten or eleven decibels, is larger than all the values of *"quantum advantage"* presented in the literature, and cancels *"ad abundantiam"* any quantum advantage in any comparison with the Noise Radar technology and with any classical radar using *"phase only"* (constant amplitude) signal coding.

## 5. Conclusions

In addition to the afore-mentioned Range problems of QR, it results that the Range measurement, *"embedded"* in all conventional radars, is a difficult issue in Quantum Radar, and the proposed solutions – out of the scope of this paper – add complexity to a yet – complicated equipment.

Other relevant considerations such as technical feasibility, operational problems and, last but not least, cost are found in [39] and [40]. Regarding the cost, QRs require costly cryogenic generators (in the $mK$ range) using Helium-4 and in some cases the hardly available Helium-3.

A good synthesis on the operational problems of QR and its readiness is presented in [61] with numerous References. In [61] the signal-to-noise (a.k.a. power budget) problem, discussed here, is not the main scope, but it was probably known to the author. The conclusion of [61] includes what follows: *"Ultimately, one should not dwell on the – black-and-white what is better – mentality, but rather pursue this technology with the mind-set of intellectual curiosity and ignorance on its most appropriate application"*.

This general approach would be acceptable, or even welcome, if applied to basic research. However, radar is an object of applied research, in which, for ethical reasons, we do not agree with promising impossible results to financing Bodies, especially when the impossibility may be shown with a few evaluations written on the back of an envelope.

The above considerations indicate that, at the present status of knowledge, there is no reason why a Quantum Radar, irrespective of the used protocol, shall perform better than a Conventional Radar. Therefore, the rationale of some claims found in [11], regarding stealth targets, remains unclear, see for instance: *"In this paper, for the first time, a microwave quantum radar setup based on quantum illumination protocol and using a Josephson Traveling Wave Parametric Amplifier (JTWPA) is proposed. … Measurement results of the developed JTWPA, pumped at 12 GHz, show the capability to generate entangled modes in the X-band, making our MQI system a promising candidate for the detection of stealth objects"*.

- With a constraint on the transmitting power, a limited quantum advantage is alleged in some literature.
- The random nature of the transmitted signal does not permit any *"tailoring"*, resulting to a PAPR-related loss significantly greater that the above advantage.
- If a low-powered signal of a quantum noise radar is amplified, then a classical noise radar results, which outperforms the quantum radar.
- If enough noise is added at the idler level, such as when it is amplified or measured heterodyne, then all the quantum advantage is lost.
- Quantum radars are more difficult to achieve than what recent experiments were claiming, and the work with signals photons in the microwave or mm-wave systems seems not to be a fruitful idea.

Conclusions similar to the ones of this paper are finally appearing in widely-distributed journals such as Science [62] from which we report the following: *"Even if experimenters can overcome the technical hurdles, quantum radar would still suffer from a fatal weakness, researchers say. The entangled pulses of microwaves provide an advantage only when the broadcast pulses are extremely faint. The extra quantum correlations fade from prominence if pulses contain significantly more than one photon—which is overwhelmingly the case in real radar. 'If you crank up the power, you won't see any difference between the quantum and the classical,' Barzanjeh says. And cranking up the power is a much easier way to improve the sensitivity"*.

Again in [62] it is noticed that it is difficult to establish a useful and practical microwave application of quantum sensing even with the full advantage by an entangled source (e.g. the promised 6 $dB$ advantage) when a simpler classical system will perform better with a higher power output and a cheaper and simpler setup. Furthermore, the alleged military advantage of a quantum radar due to its covertness, i.e. the LPI features, is practically immaterial due to its extremely short operating range.